\documentclass[conference]{IEEEtran}
\IEEEoverridecommandlockouts

\usepackage{amsmath,amsfonts}
\usepackage{algorithm}
\usepackage{array}
\usepackage[caption=false,font=normalsize,labelfont=sf,textfont=sf]{subfig}
\usepackage{textcomp}
\usepackage{stfloats}
\usepackage{url}
\usepackage{verbatim}
\usepackage{graphicx}
\usepackage{cite}
\usepackage{makecell}

\usepackage{cmap}
\usepackage[T1]{fontenc}
\usepackage{cite}
\usepackage{amsmath,amssymb,amsfonts}
\usepackage{graphicx}
\usepackage{textcomp}
\usepackage{xcolor}
\usepackage{bm}
\usepackage{amsthm}
\usepackage{enumerate}
\usepackage{booktabs}
\usepackage{diagbox}
\usepackage{supertabular}
\usepackage{subfloat}
\usepackage{caption}

\usepackage{CJK}
\usepackage{indentfirst}
\usepackage{amsmath}
\usepackage{cases}
\usepackage{mathrsfs}
\usepackage{algorithm}
\usepackage{algpseudocode}
\usepackage{makecell}
\def\BibTeX{{\rm B\kern-.05em{\sc i\kern-.025em b}\kern-.08em
    T\kern-.1667em\lower.7ex\hbox{E}\kern-.125emX}}

\begin{document}

\title{Learning-based Power Control for Secure Covert Semantic Communication}
\author{
\IEEEauthorblockN{Yansheng Liu\IEEEauthorrefmark{1}, Jinbo Wen\IEEEauthorrefmark{1}, Zongyao Zhang\IEEEauthorrefmark{1}, Kun Zhu\IEEEauthorrefmark{1}, Yang Zhang\IEEEauthorrefmark{1}, Jiangtian Nie\IEEEauthorrefmark{2}, Jiawen Kang\IEEEauthorrefmark{3}}

\IEEEauthorblockA{
\IEEEauthorrefmark{1}\textit{College of Computer Science and Technology, Nanjing University of Aeronautics and Astronautics, China}\\
\IEEEauthorrefmark{2}\textit{College of Computing and Data Science, Nanyang Technological University, Singapore}\\
\IEEEauthorrefmark{3}\textit{School of Automation, Guangdong University of Technology, China}\\
}
\IEEEcompsocitemizethanks{
\textit{Corresponding author: Kun Zhu (e-mail: zhukun@nuaa.edu.cn).}
}
}

\maketitle

\begin{abstract}
Semantic Communication (SemCom), as a next-generation communication technology, promises to enhance message delivery efficiency while reducing network resource consumption. Despite progress in SemCom, research on SemCom security is still in its infancy. To bridge this gap, we propose a general covert SemCom framework for wireless networks, which introduces the application of covert communications aided by a friendly jammer, thereby reducing the risk of eavesdropping. Our approach transmits semantic information covertly, making it difficult for wardens to detect. Given the aim of maximizing covert SemCom performance, we formulate a power control problem in covert SemCom under energy constraints. Furthermore, we propose a learning-based approach based on the soft actor-critic algorithm, optimizing the power of the transmitter and the friendly jammer. Our numerical findings substantiate the efficacy of our proposed approach in bolstering covert SemCom performance.


\end{abstract}

\begin{IEEEkeywords}
Convert SemCom, wireless communication, power control, deep reinforcement learning.
\end{IEEEkeywords}

\section{Introduction}
As a new communication paradigm, Semantic Communication (SemCom) focuses on transferring task-relevant information, instead of on bit-level transmission \cite{du2023rethinking}. The basic architecture of SemCom involves joint training of semantic encoder and decoder. After passing through the semantic encoder, the message is transformed into semantic information for wireless transmission. The semantic decoder can then decode this information to recover the source message or fulfill specific task requirements. With the lightening of semantic models, this emerging form of communication presents unprecedented opportunities for network edge devices. 

However, there are inherent challenges in SemCom security. In semantic communication systems, an attacker can target not only the transmission of semantic information, as is typical in conventional communication systems, but also the machine learning models used for semantic information extraction since most semantic data is generated through ML-based methods. Traditional communication security technologies cannot be directly applied to SemCom because they primarily focus on protecting the raw data (e.g., bits and symbols) instead of semantic information, one key reason is the lack of new security performance indicators \cite{du2023rethinking}. Moreover, highly complex encryption mechanisms impose enormous pressures on low-power network edge devices. With the development of covert communication and SemCom, covert SemCom has emerged as a novel approach to tackle the inherent security challenges in SemCom. In covert SemCom, the jammer transmits noise signals to conceal the transmitter's activity, making potential wardens even unable to confirm the occurrence of communication activities \cite{9771821}. Therefore, the communication process is hidden. And even if the communication is partially sniffed, since semantic-level encoding is adopted~\cite{yang2024secure}, the warden may not understand the true meaning of the sniffed information, thereby enhancing the concealment and security of communication under limited energy conditions.

However, a significant challenge for covert SemCom lies in improving concealment while maximizing the performance of SemCom under limited energy conditions. Current research on covert communication primarily focuses on traditional communication paradigms \cite{zhang2024research, 10495329, 10064054}. In addition, research in the domain of SemCom security is still at a nascent stage \cite{yang2024secure, Luo_Chen_Tao_Yang_2022}. To the best of our knowledge, covert SemCom has been rarely studied. The authors in \cite{wang2023multi} proposed a covert SemCom framework, which is tailored to specific image question-answering tasks. Thus, covert SemCom necessitates more comprehensive investigations.

To address the challenge mentioned above, we propose a novel general covert SemCom framework to tackle the security challenge of SemCom under limited energy conditions. Specifically, we consider the power control problem in covert SemCom to ensure semantic communication quality under overall energy limitations. Then, we propose a learning-based approach based on the Soft Actor-Critic (SAC) algorithm to optimize power control within SemCom systems. Our contributions are summarized as follows:
\begin{itemize}
\item[$\bullet$] \textit{A general covert semantic system framework}: The proposed framework seamlessly integrates covert communication principles, enabling its application across diverse data modalities including text, image, and audio. This ensures its versatility and practicality in the context of multimodal data processing. Irrespective of the data type involved, our framework facilitates efficient and concealed semantic transmission.
\item[$\bullet$] \textit{Power control for covert semantic communication}: We consider power control for the covert SemCom scenario composed of a transmitter, a friendly jammer, a receiver, a warden, and a power regulator. Considering energy constraints and the aim of maximizing the performance of covert SemCom, we strive to optimize the power of the transmitter and the jammer, thus enhancing covert SemCom performance under limited energy conditions.
\item[$\bullet$] \textit{Learning-based approach for covert semantic communication power control}: We propose a SAC-based approach to power control in covert semantic communication, which not only safeguards the efficiency of covert communication but also accomplishes high-quality semantic decoding. Numerical results demonstrate that the proposed approach outperforms other deep reinforcement learning algorithms.
\end{itemize}

\section{System Model and Problem Formulation}
\subsection{System Model}
\begin{figure}[!t]
    \centering  
    \includegraphics[width=0.39\textwidth]{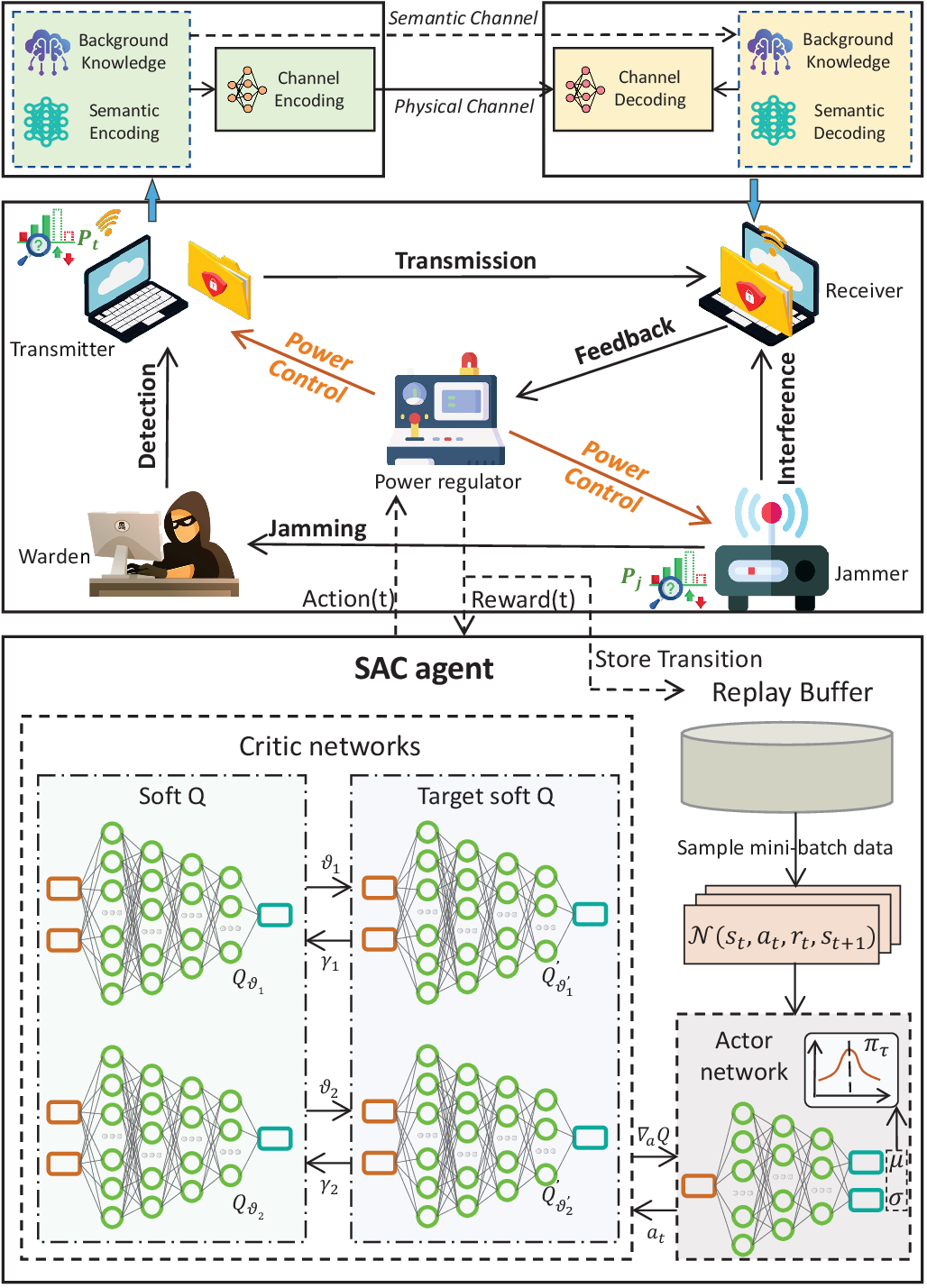}   
    \captionsetup{font=footnotesize}
    \caption{An illustration of learning-based power control for secure covert semantic communication.}
    \label{fig:system}
\end{figure}

The proposed framework for covert SemCom is depicted in Fig.~\ref{fig:system}. The communication scenario is composed of five components: a transmitter, a receiver, a friendly jammer, a warden, and a power regulator, all of which operate within an open wireless environment. For simplicity, we take text SemCom as an example, the primary objective of the network system is to enable the transmitter to send extracted textual semantic information to the receiver without being detected by the warden, ensuring that the receiver can successfully comprehend the semantic information. The transmitter consists of a semantic encoder that extracts semantic features from the text to be transmitted and a channel encoder that generates symbols to facilitate subsequent transmission. The receiver is equipped with a channel decoder for symbol detection and a semantic decoder for text estimation \cite{Xie_Qin_Li_Juang_2021}. 

To enhance communication security, we introduce the application of covert communications by a friendly jammer and a power regulator into the environment, which optimizes power control between the transmitter and the jammer. The role of the jammer is to inject noise that interferes with any potential warden to intercept the communication, thereby hindering unauthorized access to the information being transmitted. In addition, the role of the power regulator is to formulate and enact power strategies, gather feedback from the receiver, and persistently fine-tune these strategies for optimum performance. Unlike other physical layer security methods, covert communication improves the security of information by hiding the act of communication \cite{10570800}. Specifically, the warden employs statistical methods to differentiate the transmitter's activity status. Accordingly, the warden assesses two potential scenarios by collecting signal samples in the channel. The null hypothesis, $\mathcal{H}_0$, indicates that the transmitter is inactive, but the jammer is active. The signal power detected by the warden can be mathematically represented as \cite{du2024generative}
\begin{equation}
\begin{split}
    P_{{\text{eval}}_{\mathcal{H}_0}} = P_j  \left(\frac{\lambda}{4\pi D_{jw}}\right)^{\alpha_{jw}}  \left|h_{jw}\right|^2 + \mathcal{K}^2.
\end{split}
\end{equation}

Considering the alternative hypothesis, represented as $\mathcal{H}_1$, this hypothesis suggests that the transmitter and the jammer are all active. Under this hypothesis, the signal power detected by the warden can be mathematically represented as \cite{du2024generative}
\begin{equation}
    \begin{split}
    P_{\text{eval}_{\mathcal{H}_1}} = &\mathcal{K}^2 + P_t \left(\frac{\lambda}{4\pi D_{tw}}\right)^{\alpha_{tw}}  |h_{tw}|^2 \\ 
    &+ P_j  \left(\frac{\lambda}{4\pi D_{jw}}\right)^{\alpha_{jw}}  |h_{jw}|^2,
    \end{split}
\end{equation}
where $P_t$ is the transmit power, $P_j$ is the jamming power, $\mathcal{K}^2$ is the Gaussian noise, and $\lambda$ is the wavelength of the carrier signal. The distances between the jammer and the warden, and the transmitter and the warden are given by $D_{jw}$ and $D_{tw}$, respectively. The path loss exponents $\alpha_{jw}$ and $\alpha_{tw}$ account for the attenuation of the signal as it propagates through the medium for their respective links. The small-scale fading effects are captured by $h_{jw}$ and $h_{tw}$, representing the multipath propagation effects. The wavelength $\lambda$ can be derived from the frequency of the signal as $\lambda = \frac{c}{f}$, where $c$ is the speed of light in a vacuum and $f$ is the frequency of the transmitted signal.

Based on the aforementioned hypotheses, the eavesdropping decision-making of the warden is denoted as $\mathcal{D}_0$ for the null hypothesis $\mathcal{H}_0$ and $\mathcal{D}_1$ for the alternative hypothesis $\mathcal{H}_1$, following a threshold rule \cite{chen2023covert}. The warden employs statistical hypothesis testing methods to evaluate the transmitter's activity as either $\mathcal{H}_0$ or $\mathcal{H}_1$. Specifically, there are two types of detection failures for the warden: a false alarm, where the decision is $\mathcal{D}_1$ during $\mathcal{H}_0$, and a miss detection, where the decision is $\mathcal{D}_0$ during $\mathcal{H}_1$. 
\begin{itemize}
    \item \textit{False Alarm:} This occurs under hypothesis \( \mathcal{H}_0 \) when there is no signal, but the warden incorrectly decides \( \mathcal{D}_1 \), indicating the presence of a signal. 
    \item \textit{Miss Detection:} This occurs under hypothesis \( \mathcal{H}_1 \) when a signal is present, but the warden incorrectly decides \( \mathcal{D}_0 \), indicating no signal. 
\end{itemize}

Combine the above two cases, the Detection Error Probability (DEP) can be characterized as
\begin{equation}
\begin{split}
\label{dep}
    \mathbb{P}_{\text{DEP}} = \mathbb{P}_{\text{FA}} + \mathbb{P}_{\text{MD}}= \mathbb{P}(P_{\text{eval}_{\mathcal{H}_0}} > \tau) + \mathbb{P}(P_{\text{eval}_{\mathcal{H}_1}} < \tau),
\end{split}
\end{equation}
where \( \tau \) denotes the detection threshold, $\mathbb{P}_{\text{FA}}$ and $\mathbb{P}_{\text{MD}}$ denote the probability of false alarm and the probability of miss detection, respectively. 



\subsection{Problem Formulation}
\label{sec:prob}
For simplicity, taking text SemCom as an example, we employ the SemCom system based on the transformer architecture designed in \cite{Xie_Qin_Li_Juang_2021}. In this scheme, the transmitter processes the input raw data, referred to as \textit{$Info_s$}. Upon receiving the semantic information, the receiver utilizes a corresponding decoder to reconstruct the original informational content, hereafter referred to as \textit{$Info_r$}.

During the channel transmission process, the Signal to Interference plus Noise Ratio (SINR) is influenced by the power of the transmitter and the jammer, directly altering the Bit Error Rate (BER) and consequently impacting the receiver's information reconstruction accuracy. For text transmission, BER does not accurately reflect semantic performance. In addition to human judgment to assess the similarity between sentences, the Bilingual Evaluation Understudy (BLEU) score is commonly used to measure the results in machine translation \cite{belghazi2018mutual}. For the transmitted sentence \textit{$Info_s$} with length $l_s$ and the decoded sentence \textit{$Info_r$} with $l_r$, the BLEU can be expressed as \cite{Xie_Qin_Li_Juang_2021}
\begin{equation}
\begin{aligned}
\log \text{BLEU} = \min\left(1 - \frac{l_r}{l_s}, 0\right) + \sum_{n=1}^{N} w_n\log s_n,
\end{aligned}
\end{equation}
where $w_n$ is the weight of $n$-grams and $s_n$ is the $n$-grams score, as given by
\begin{equation}
\begin{aligned}
s_n = \frac{\sum_{k} \min(C_k(Info_r), C_k(Info_s))}{\sum_{k} \min(C_k(Info_r))},
\end{aligned}
\end{equation}
where $C_k(\cdot)$ is the frequency count function for the $k$-th elements in $n$-th grams.

We adopt the BLEU metric as our objective function to assess the semantic retention and reconstruction quality of the information. The energy factors \( \eta_t \) and \( \eta_j \) signify the energetic costs per unit for transmission power and jamming power, respectively. We formulate the optimization problem as 
\begin{subequations}
\begin{align}
\label{eq:opt}
    & \max_{P_t, P_j} \quad \text{BLEU} (Info_{\text{s}}, Info_{\text{r}}) \\
    & \quad \rm{s.t.} \quad\:\:\:\mathbb{P}_{\text{DEP}} > \zeta_{\text{th}}, \label{eq:opta}\\
    & \qquad \qquad \eta_t P_t \leq E_t, \label{eq:optb} \\
    & \qquad \qquad \eta_j P_j \leq E_j, \label{eq:optc}
\end{align}
\end{subequations}
where the first constraint~(\ref{eq:opta}) is the covert communication constraint with a utility threshold $\zeta_{\text{th}}$. Covert communication is considered successful when DEP exceeds a certain threshold \( \zeta_{\text{th}} \), which approximates to 1 \cite{du2024generative}. The second and third constraints~(\ref{eq:optb}) and (\ref{eq:optc}) are the energy constraints of the transmitter and the jammer, respectively. 

Our optimization problem aims to maximize the semantic performance (i.e., BLEU score) while meeting the total energy constraint, and ensuring that the communications remain the covert communication constraint. Under the condition of limited energy $E_t$ and $E_j$, it is obvious that a lower value of transmit power $P_t$ or a higher value of jamming power $P_j$ facilitates easier maintenance of the covert communication constraint. Specifically, when $P_t$ is too low, $\mathbb{P}_{\text{MD}}$ in (\ref{dep}) easily approaches or even exceeds the certain threshold $\zeta_{\text{th}}$, so as to satisfy the covert communication constraint. Similarly, when $P_j$ is too high, $\mathbb{P}_{\text{FA}}$ in (\ref{dep}) also tends to approach or even exceed the certain threshold $\zeta_{\text{th}}$, thereby satisfying the covert communication constraint. 

However, this setting tends to increase the BEP due to a reduced SINR, which in turn diminishes the BLEU score. As such, joint optimization is essential to balance covert communication and the quality of the regenerated information. Moreover, it has to be said that the aforementioned problem is complicated because the semantic metric BLEU, $P_t$, and $P_j$ are not an explicit mathematical relationship, rendering it challenging to solve using traditional mathematical techniques. Deep reinforcement learning, particularly the SAC, is effective for dynamic and high-dimensional environments, as it adjusts to environmental changes and ensures efficient learning in large problem spaces. The principle of maximum entropy allows it to fully explore during decision-making, avoiding local optima and enhancing the robustness of learning \cite{mock2023comparison}. Therefore, we adopt SAC for optimal power control, giving the solution to the optimization problem in Section~\ref{sec:gdm}.

\section{SAC-based Optimal Power Control}
\label{sec:gdm}
We elaborate on the utilization of the SAC algorithm for optimal power control. The algorithm provides a power control scheme, typically denoted as $\{P_t, P_j\}$, as seen in (\ref{eq:opt}). The core variables used in this paper and their descriptions are compiled in Table~\ref{tab:variables} for easy reference. The reward of the agent (i.e., the power regulator) is determined by the actions executed pertaining to a particular state of the environment. Here are the details of the SAC formulation:
\begin{itemize}
    \item \textit{State space:} The states, denoted as $\mathbf{s}$, are construed as a vector of measurements that encapsulate the conditions of the environment. In alignment with our discussion in Section~\ref{sec:prob}, herein, the environmental vector refers to the comprehensive set of all variables influencing the optimal power control scheme, given by
    \begin{equation}
    \begin{aligned}
    \mathbf{s} = \{D_{\text{tw}}, D_{\text{tr}}, D_{\text{jw}}, D_{\text{jr}}, \alpha_{\text{tw}}, \alpha_{\text{tr}}, \alpha_{\text{jw}}, \alpha_{\text{jr}}, \\ \zeta_{\text{th}}, h_{\text{tw}}, h_{\text{tr}}, h_{\text{jw}}, h_{\text{jr}}, \mathcal{K}^2, \lambda, \tau\}.
    \end{aligned}
    \end{equation}

    \item \textit{Action space:} Based on the environment state, the agent initiates an action, denoted as $\mathbf{a}$. The action space represents the control actions taken by the power regulator, structured with several control variables. The proposed scheme uses transmit power $P_t$ and jamming power $P_j$ as actions of the power regulator, given by
    \begin{equation}
    \begin{aligned}
    \mathbf{a} = \{P_t, P_j\}.
    \end{aligned}
    \end{equation}
    \item \textit{Immediate reward:} In optimal power control, the goal is to maximize the BLEU score while minimizing the total energy consumption given by \( \eta_t \times P_t \), \( \eta_j \times P_j \), and ensuring the communications remain the covert communication constraint. To speed up the convergence, we consider the above two constraints in designing the reward functions, which is expressed as
    \begin{equation}
    \begin{aligned}
    \label{eq:reward_function}
    R{(t)}= 
    \begin{cases}
    \text{BLEU}{(t)}, & \text{\makecell*[l]{if (\ref{eq:opta}), (\ref{eq:optb}), and (\ref{eq:optc}) are met}},  \\
    \Psi, & \text{otherwise,}
    \end{cases}
    \end{aligned}
    \end{equation}
    where $R{(t)}$ is the reward at time $t$, $\text{BLEU}{(t)}$ represents the degree of semantic retention and reconstruction quality of the source data and reconstructed data at time $t$, and $\Psi$ represents the penalty factor, which varies with the scale of the environment. 
    \item \textit{Value function:} SAC meticulously optimizes a stochastic policy, maximizing both long-term entropy and expected lifetime rewards. Concomitantly, it learns a policy $\pi_{\tau} (\mathbf{a}|\mathbf{s})$ and two Q-functions, $Q_{\theta_1}$ and $Q_{\theta_2}$, applying the minimum of these two Q-values to construct the targets in Bellman error functions, which is given by
    \begin{equation}
    y(r, \mathbf{s}, d) = \gamma (1 - d) \big( \min_{i = 1,2} Q_{\theta_i}(\mathbf{s}, \mathbf{a}) - \alpha \log \pi_{\tau} (\mathbf{a}|\mathbf{s}) \big)  + r,
    \label{eq:bellman}
    \end{equation}
    where $d$ is indicative of the done signal, $\alpha$ serves as a coefficient that manages trade-offs, $\gamma$ is the discount factor, and $r$ denotes the reward. In every state, the policy executes actions to maximize the expected future return along with the expected future entropy, i.e., optimizing the state value function $V^{\pi}(\mathbf{s})$, which is given by 
    \begin{equation}
    \begin{aligned}
    V^{\pi}(\mathbf{s}) = \mathbb{E}_{\mathbf{a} \sim \pi} \left[ Q^{\pi_{\tau}}(\mathbf{s}, \mathbf{a}) - \alpha \log \pi_{\tau}(\mathbf{a}|\mathbf{s}) \right].
    \label{eq:stateV}
    \end{aligned}
    \end{equation}
    
    \item \textit{SAC algorithm design:} We consider the power regulator as a SAC agent and a stochastic policy $\pi(\mathbf{a}_{n}|\mathbf{s}_{n})$ is defined to map from states to a probability distribution over actions. At each training iteration, the power regulator observes the state $\mathbf{s}_n$ and executes an action $\mathbf{a}_n$ sampled from the current policy $\pi$. Then, the environment turns into the next state $\mathbf{s}_{n+1}$ and sends back the reward $r_n$ to the power regulator. The tuple $\mathbf{e}_n$ is stored as experience in a data set $\mathcal{D}$ called replay buffer. By sampling from $\mathcal{D}$ and updating the policy periodically according to a learning algorithm, the power regulator will finally find an optimal policy $\pi^*$ that maximizes the long-term reward (\ref{eq:rew}). The specific learning process can be seen from the Algorithm \ref{alg:SAC}.

    \begin{equation}
    \begin{aligned}
    \label{eq:rew}
    R_{\pi} = \mathbb{E}_\pi \left[ \sum_{n=0}^{\infty} \gamma^n r(\mathbf{s}_n, \mathbf{a}_n) \right].
    \end{aligned}
    \end{equation}
    
\end{itemize}

The computational complexity of the proposed SAC algorithm in Fig.~\ref{fig:system} incorporates the parameters updating of three neural networks i.e., \(Q_{\theta_1}\), \(Q_{{\theta_2}}\), and \(\pi_{\tau}\). Therefore, the computation of the complexity of Algorithm~\ref{alg:SAC} is
$\mathcal{O}\left(\sum_{j=0}^{J-1} n_{j}^{Q} n_{j+1}^{Q} + \sum_{k=0}^{K-1} n_{k}^{\pi} n_{k+1}^{\pi}\right)$, where ${J}$ symbolizes the count of fully connected layers in the $Q_{\theta_1}$ and $Q_{\theta_2}$ networks, both of which have identical structure, and ${K}$ denotes that for $\pi_\tau$ network. $n_{j}^{Q}$ and $n_{k}^{\pi}$ denote the count of neurons at the ${j}$-th layer of ${Q}_{\theta_1}$ or ${Q}_{\theta_2}$ networks and the ${k}$-th layer of ${\pi}_{\tau}$ network, respectively. ${j}$ = 0 and ${k}$ = 0 represent the input layers, respectively.


\addtolength{\topmargin}{0.05in}
\begin{algorithm}[!t]
\caption{SAC-based Power Control in Covert SemCom}
\label{alg:SAC}
\begin{algorithmic}[1]
\State \textbf{Input:} States from the environment $\mathbf{s}$;
\State \textbf{Output:} Actions to the environment $\mathbf{a} = \{P_t, P_j, S\}$;
\State Initialize policy network parameters $\tau$, Q-function parameters $\theta_1$ and $\theta_2$;
\State Initialize the target network parameters equal primary parameters as $\hat{\theta}_1 \leftarrow \theta_1$, $\hat{\theta}_2 \leftarrow \theta_2$;
\For {episodes $e = 1, \ldots, N$}
    \State Reset environment state $s_0$ and replay buffer $\mathcal{D}$;
    \For {time-slot $t = 1, \ldots, T$}
        \State Take action $\mathbf{a}(t)$ based on $\pi_{\tau} (.|\mathbf{s})$;
        \State Execute the control actions to the environment;
        \State Calculate rewards through (\ref{eq:reward_function});
        \State Update $\mathbf{s}_{t}$ into $\mathbf{s}_{t+1}$, and store transition into $\mathcal{D}$;
        \State Sample a random mini-batch of data $\mathcal{B}$ with a size \textit{N} from $\mathcal{D}$;
        \State Update the target Q functions using (\ref{eq:bellman});
        \State Update the Q function as
        \Statex $\Delta \theta_i = \frac{1}{|N|} \sum_{(\mathbf{s},\mathbf{a},r,\mathbf{s'}) \in \mathcal{B}} (Q_{\theta_i} (\mathbf{s}, \mathbf{a}) - y(r, \mathbf{s'}, d))^2$, for $i = 1, 2$;
        \State Update the target network as 
        \Statex $\hat{\theta}_i \leftarrow \beta \hat{\theta}_i + (1 - \beta) \theta_i$, for $i = 1, 2$;
        \State Update the policy network as
        \Statex $\Delta \tau = \frac{1}{|N|} \sum_{\mathbf{s} \in \mathcal{B}} (Q_{\theta_i} (\mathbf{s}, a_{\tau} (\mathbf{s})) - \alpha \log \pi_{\tau} (a_{\tau} (\mathbf{s})|\mathbf{s}) - y(r, \mathbf{s'}, d))^2$, for $i = 1, 2$;
    \EndFor        
    \If {reach maximum episodes \textit{N}}
        \State Break;
    \EndIf
\EndFor
\end{algorithmic}
\end{algorithm}
\section{Numerical Results}
\begin{table}[t]
	\renewcommand{\arraystretch}{1.1}
        \captionsetup{font = small}
	\caption{Description of Key Hyperparameters. }\label{tab:variables} \centering 
	\begin{tabular}{m{5.3cm}<{\raggedright}|m{2.1cm}<{\centering}}	 	
        \toprule[1.5pt]
		\hline		
		\textbf{Descriptions} & \textbf{Symbols}\\	
		\hline
		Distances between the two parties, respectively.  &  $D_{\text{tw}}, D_{\text{tr}}, D_{\text{jw}}, D_{\text{jr}}$\\	
		\hline
        Path loss exponents between the two parties, respectively. &  $\alpha_{\text{tw}}, \alpha_{\text{tr}}, \alpha_{\text{jw}}, \alpha_{\text{jr}}$  \\
        \hline
		Factors reflect the small-scale fading effects between the two parties, respectively. & $h_{\text{tw}}, h_{\text{tr}}, h_{\text{jw}}, h_{\text{jr}}$ \\	
		\hline
        Energy factors signify the energetic costs per unit for transmission power and jamming power, respectively. & $\eta_t, \eta_j$ \\ 
		\hline
		Gaussian noise. &  $\mathcal{K}^2$  \\	
		\hline		
		Speed of light in a vacuum. &  $c$ \\	
		\hline		
		Frequency of the transmitted signal. &  $f$ \\
		\hline
        Wavelength of the carrier signal. & $\lambda$\\
        \hline 
        Detection threshold. & $\tau$ \\
        \hline
        Penalty factor. & $\Psi$ \\
        \hline
        \bottomrule[1.5pt]
	\end{tabular}\label{table_parameter}
\end{table}

The effects of increasing transmit power on various parameters, such as the covert rate, defined as the data rate attainable during covert communications, DEP, and BEP, are illustrated. These effects are observed when the jamming power, $P_j$, is set to $45\:  \rm{dBW}$ and $\tau$ is $50$. Within a Cartesian coordinate grid where the units are in meters, the transmitter, warden, receiver, jammer, and power regulator are positioned at coordinates $(0, 0)$, $(0, 100)$, $(100, 0)$, $(100, 100)$, and $(50, 50)$, respectively. The penalty factor $\Psi$ is set to $-5$. The path loss exponents are set to $\alpha_{tr} = 1$, $\alpha_{tw} = 1.2$, and both $\alpha_{jw}$ and $\alpha_{jr}$ are configured to $1.4$. As for small-scale channel fading attributes such as $h_{tw}$, $h_{tr}$ , $h_{jw}$, and $h_{jr}$, they are designed to conform with the $\alpha-\mu$ fading model, with the parameters $\alpha$ and $\mu$ being $2$ and $4$, respectively \cite{yacoub2007alpha}. 

\begin{figure}[!t]
    \centering  
    \includegraphics[width=0.4\textwidth]{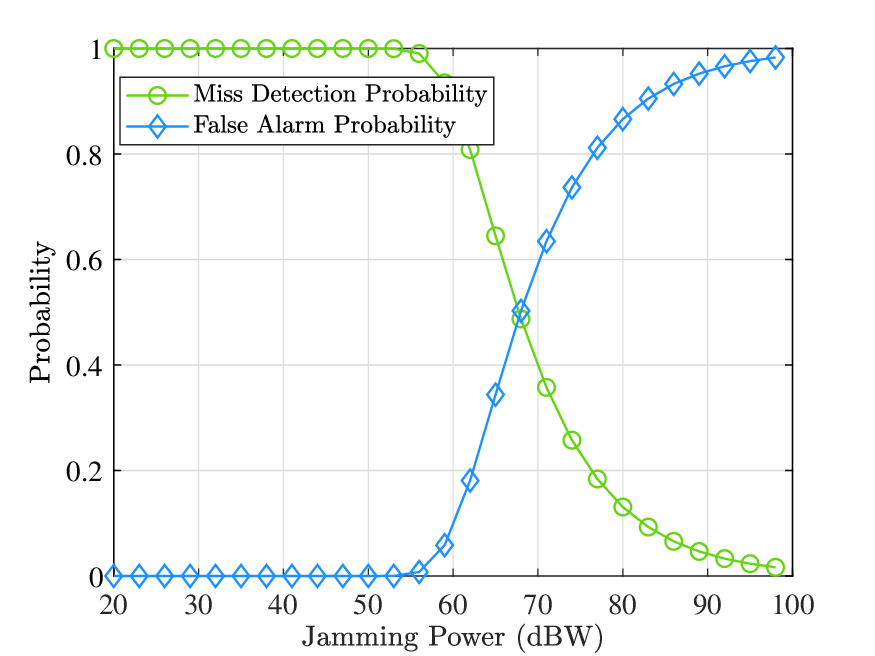}   
    \captionsetup{font=footnotesize}
    \caption{The False Alarm Probability $\mathbb{P}_{\text{FA}}$ and the Miss Detection Probability $\mathbb{P}_{\text{MD}}$ versus the jamming power, and the transmit power $P_t$ is $45\:\rm{dBW}$.}
    \label{fig:fig44}
\end{figure}

\begin{figure}[!t]
    \centering  
    \includegraphics[width=0.4\textwidth]{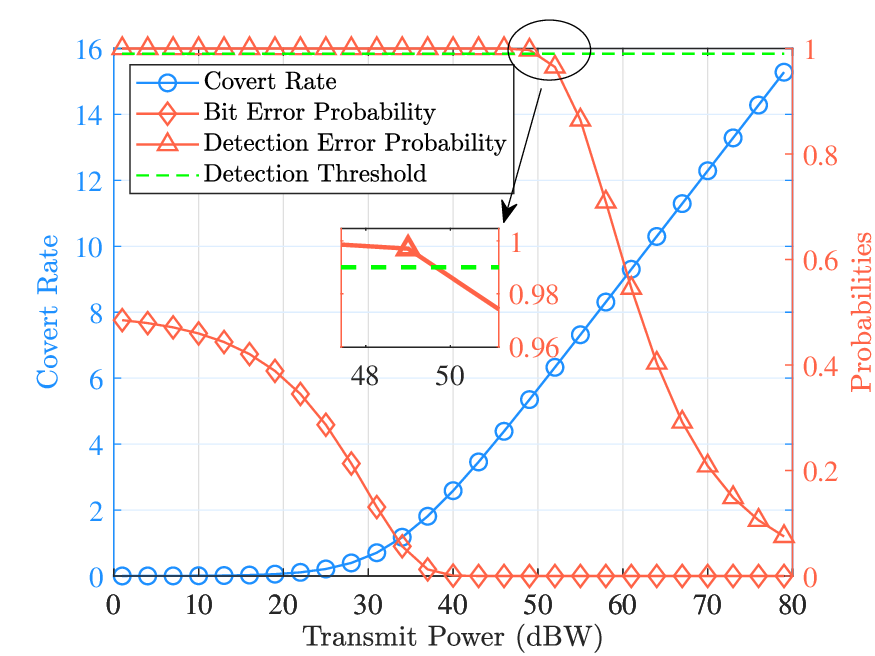}   
    \captionsetup{font=footnotesize}
    \caption{The covert rate, DEP, and the BEP versus the transmit power, and the jamming power $P_j$ is $45\:\rm{dBW}$.}
    \label{fig:fig1}
\end{figure}

The adopted channel coding strategy is Binary Phase-shift Keying (BPSK). Fig.~\ref{fig:fig44} illustrates the relationship between the false alarm probability and the missed detection probability with respect to interference power, given the transmission power. It is evident that the results are consistent with the analysis above. Moreover, as demonstrated in Fig.~\ref{fig:fig1}, given the jamming power, the gradual increase in transmission power undoubtedly leads to a direct increase in the covert rate. However, as the transmission power gradually increases, the signal power received by the warden gradually reaches its detection range. This increases the probability of being monitored by the warden, i.e., the DEP of the warden gradually decreases. For the receiver, there are not only data signals from the transmitter in the environment but also noise signals from the friendly jammer. When the transmit power is smaller and the noise power is larger, even though the trained SemCom model can well extract the semantic information, the final recovery result will be very poor due to the high BEP. In addition, when the transmission power is approximately $49.5$ $\rm{dBW}$, it can ensure covert communication while creating good channel conditions for semantic information reconstruction.

\begin{figure}[!t]
    \centering  
    \includegraphics[width=0.4\textwidth]{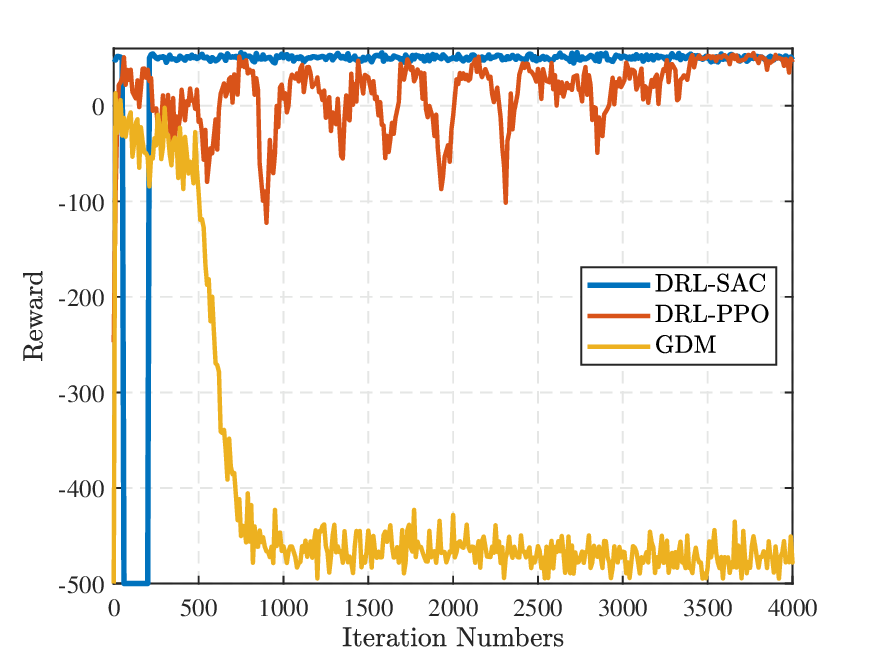}   
    \captionsetup{font=footnotesize}
    \caption{Test rewards of the proposed SAC-based algorithm and other algorithms, with batch size ${N} = 512$, discount factor $\gamma = 0.95$, and the learning rate of actor and critic networks is $10^{-4}$.}
    \label{fig:fig2}
\end{figure}
\begin{figure}[!t]
    \centering  
    \includegraphics[width=0.39\textwidth]{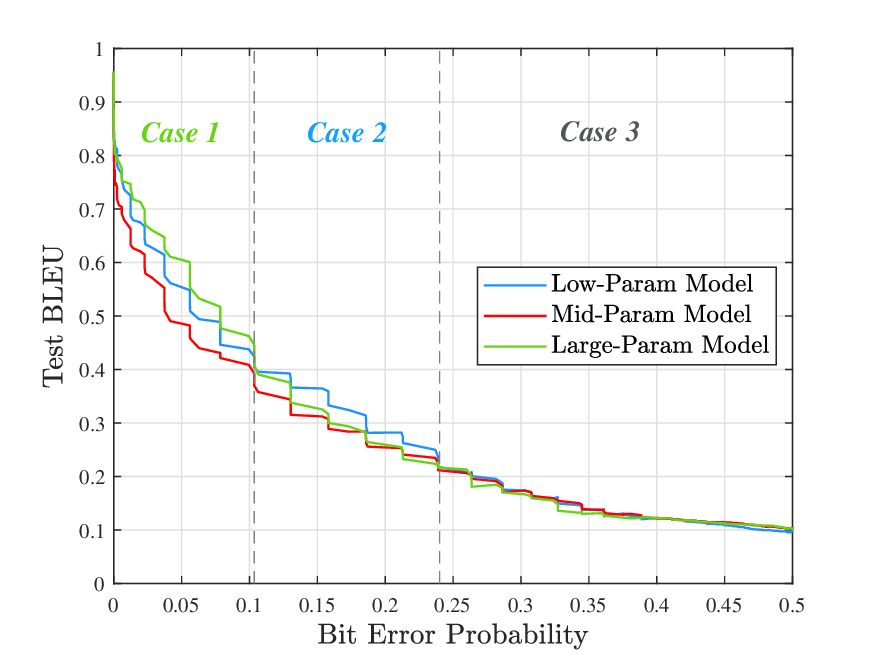}   
    \captionsetup{font=footnotesize}
    \caption{The quality of data reconstruction when the receiver with different scales of the model under different BEP conditions.}
    \label{fig:fig3}
\end{figure}

Fig.~\ref{fig:fig2} presents the performance analysis of the proposed SAC-based algorithm for optimal power control. We compare the proposed SAC-based algorithm with other algorithms: 1) \textit{PPO-based algorithm} that the power regulator utilizes prioritizes optimizing control strategies based on data reconstruction performance; 2) \textit{GDM-based algorithm} that the power regulator designs schemes by denoising noise to the initial Gaussian noise to recover or discover the optimal actions \cite{wen2024generative}. As shown in Fig.~\ref{fig:fig2}, the proposed SAC-based algorithm outperforms PPO and GDM algorithms. Specifically, the SAC-based algorithm converges to good results faster compared to PPO, while PPO achieves satisfactory results after approximately 3500 iterations, indicating that the proposed SAC-based can more effectively find the optimal scheme.  The reason is that the SAC algorithm, by maximizing policy entropy, helps converge to the optimal solution more effectively, enabling it to reach better solutions in a shorter time compared to the PPO and GDM algorithms \cite{mock2023comparison, wen2024generative}. 

Fig.~\ref{fig:fig3} shows the reconstruction process under varying BEPs. We can observe that when BEP is relatively small in \textbf{\textit{Case 1}}, the large-scale models significantly outperform the other two. This is because the transmission power is higher than the noise, and under sufficiently good channel conditions, large-scale models can effectively extract and reconstruct semantic information. However, as BEP gradually increases from $0.11$ to $0.24$ in \textbf{\textit{Case 2}}, it is clear that small-scale models can achieve better performance. This is due to the low transmission power or high interference power, resulting in a low signal-to-noise ratio. The small-scale models can alleviate the interference caused by noise more effectively compared with the large-scale models. Finally, when BEP exceeds $0.24$ in \textbf{\textit{Case 3}}, the channel conditions have already become very poor and any type of model is unable to decode semantic information correctly. The reconstructed data always appear as noise, resulting in relatively low BLEU scores. This further emphasizes the importance of optimizing power control.

\section{Conclusion}
In this paper, we have proposed a learning-based approach to optimal power control for the proposed covert SemCom, aimed at resolving the security dilemmas inherent in SemCom. Specifically, we have introduced a power regulator and a friendly jammer to achieve covert communication within the context of SemCom. To address the increased energy usage introduced by covert SemCom, we have formulated power control as an optimization problem and utilized the SAC algorithm to optimize power control between the transmitter and the friendly jammer, which strikes a balance between the effectiveness of covert transmission and SemCom. Numerical results have demonstrated the effectiveness of the proposed optimization strategy within the covert SemCom framework. For future work, we will explore ways to integrate the concept of the Mixture of Experts (MoE) system into our covert SemCom framework, ensuring reliable communication across varied network conditions.

\bibliographystyle{IEEEtran}
\bibliography{ref}

\end{document}